\documentclass[%
 reprint,
 amsmath,amssymb,
 aps,
]{revtex4-2}

\usepackage{graphicx}
\usepackage{epstopdf}
\usepackage{dcolumn}
\usepackage{bm}
\usepackage{color}
\usepackage[mathlines]{lineno}
\usepackage[breaklinks=true,colorlinks=true,linkcolor=blue,urlcolor=blue,citecolor=blue]{hyperref}
\usepackage{ragged2e}
\usepackage{enumitem}
\usepackage{bibunits}
\usepackage{amsthm}
\usepackage{diagbox}
\usepackage{multirow}
\usepackage{tabularx}

\begin{document}

\preprint{APS/123-QED}

\title{Asymmetric interaction preference induces cooperation in human-agent hybrid game}

\author{Danyang Jia$^1$}
\author{Xiangfeng Dai$^{2}$}
\author{Junliang Xing$^{1}$}
\email{jlxing@tsinghua.edu.cn}
\author{Pin Tao$^{1}$}
\author{Yuanchun Shi$^{1}$}
\author{Zhen Wang$^3$}
\email{zhenwang0@gmail.com}
\affiliation{
\vspace{2mm}
\mbox{1. Department of Computer Science and Technology, Tsinghua University, Beijing, 100084 China}
\mbox{2. School of Mechanical Engineering, Northwestern Polytechnical University, Xi'an, 710072 China}
\mbox{3. School of Cybersecurity, Northwestern Polytechnical University, Xi'an, 710072 China}
}


\begin{abstract}
With the development of artificial intelligence, human beings are increasingly interested in human-agent collaboration, which generates a series of problems about the relationship between agents and humans, such as trust and cooperation. This inevitably induces the inherent human characteristic that there are subjective interaction preferences for different groups, especially in human-agent hybrid systems where human-human interaction, agent-agent interaction, and human-agent interaction coexist. However, understanding how individual interaction preferences affect the cooperation of the system remains a major challenge. Therefore, this paper proposes a human-agent hybrid prisoner’s dilemma game system under the framework of evolutionary game. In spatial networks, the most significant difference between agents and humans is the flexibility of decision, where humans have higher adaptive capabilities, follow link dynamics, and adopt free decision rules, which enable them to choose different strategies for different neighbors. However, agents follow node dynamics and adopt consistent decision rules, applying the same strategy to different neighbors. We give the subjective preferences of any individual to different groups, involving the interaction preferences between homogeneous groups and heterogeneous groups respectively. The simulation results show that both human and agent have asymmetric interaction preferences for groups with different identities, which can significantly improve the cooperative behavior of the system. In the hybrid system, human groups show more stable prosocial behavior. Agent groups can form highly cooperative clusters under the condition of strong interaction preference for human groups. In addition, giving agents the ability to identify opponents can effectively alleviate the interaction dilemma of agents.
\begin{description}
\item[Key words]Cooperation, Human-agent game, Preference, Freedom of choice, Evolutionary game theory.
\end{description}
\end{abstract}

\maketitle


\section{Introduction}
Facing the intricate social relationships, individuals inevitably engage in fierce competition in order to obtain higher benefits. However, the development of ecosystems, economic systems and human societies cannot be achieved without cooperation, such as various countries and institutions work together to solve global climate change~\cite{keohane2016cooperation,meckling2020evolution}, develop and administer vaccines to prevent and treat diseases~\cite{kabir2019behavioral,korn2020vaccination}, as well as alleviate social conflicts between different groups~\cite{gachter2010culture,de2020group}, etc. In essence, cooperation means that the individual bears the cost of the benefit of others. In numerous dilemma games, whether it is a two-person or multi-person game, free-riding behavior is usually the optimal choice of an individual~\cite{mcnamara2004variation,kummerli2007human,santos2008social,heitzig2011self,burton2013prosocial}. Despite extensive researches carried out based on evolutionary game framework in recent decades from different perspectives~\cite{weibull1997evolutionary,smith1979game,nowak1992evolutionary}, how cooperative behavior emerges and evolves remains a mystery in many domains~\cite{cardinot2019mobility}. Especially with the development of artificial intelligence, there are more and more scenarios of human-agent integration~\cite{dafoe2021cooperative}. In order to complete tasks more efficiently and intelligently, interaction and cooperation with humans are overarching aspirations of artificial intelligence research. As the functions of agents continue to expand and diversify, a series of problems arise in the relationship between agents and humans~\cite{oksanen2020trust,mota2016playing,schniter2020trust,han2021or}, which bring greater challenges to exploring the evolution of cooperation in human-agent hybrid systems.

\subsection{Motivations and related work}
 In the development of human-agent hybrid system, most fruitful results focus on how to improve the machine intelligence, as well as the credibility of the agent intelligence, for example, by improving the ability of agent in state perception, real-time analysis, optimal decision-making, precise execution, and so on. However, it is worth emphasizing that machine intelligence does not directly measure the ability of agents to cooperate with humans, which involves more complex interaction and cooperation mechanisms that need to consider factors such as human cognition, emotion, and social behaviors in hybrid systems, so as to explore the cooperation between agents and humans more urgently. Given that in a complex system, collective behavior arises from local interactions among individuals, and a large number of studies have shown that simply changing the local interaction dynamics is enough to make the system present complex and interesting dynamic phenomena at the macro perspective. Therefore, scholars have conducted a series of fruitful research on the interactive relationship between humans and agents, including trust perception~\cite{glikson2020human,gebru2022review,mou2020would}, information exchange~\cite{burgoon2000interactivity,unhelkar2020decision}, and mental state understanding~\cite{baker2017rational,reddy2018you,buehler2020theory}, etc. Specifically, agents are endowed with anthropomorphic characteristics, such as human-like appearance, communication mode, and emotional expressions~\cite{boone2003emotional,de2016almost,christoforakos2021can}, to increase the similarity between agents and humans, thereby improving human trust in agents. Further, scholars adopted behavioral game theory to attempt to approximately measure trust and trustworthiness through empirical evidence~\cite{upadhyaya2023bot}. In the process of information exchange between humans and agents, in order to avoid cumbersome information and reduce the interference of useless information on partners' decision-making, the theory of mind based on communication is proposed, which driving the agents to decide whether, when and what information to share with humans~\cite{buehler2020theory}. There is no doubt that existing algorithms and experimental studies have demonstrated significant advances in the accuracy, robustness, and efficiency of agents in performing tasks~\cite{duan2023event}, with the aim of increasing human trust in agents. Thus, it lays a technical foundation for agents to cooperate with human better.

In addition, how agents influence the cooperative behavior of humans in conflict environments is also one of the keys to understanding the emergence of cooperation in hybrid system. Recently, some scholars tried to introduce committed agents in one-shot and anonymous social dilemma games, where committed agents always choose cooperation (or defection, etc.) and never change strategies in the absence of reputation influence and experience~\cite{guo2023facilitating,sharma2023small,shen2024prosocial}, so as to observe the evolution of prosocial behavior in normal humans. These studies reveal that for two typical population settings, well-mixed population and regular lattice population, cooperative agents have a limited effect in the prisoner’s dilemma game, but can play an important role in the stag hunt game~\cite{guo2023facilitating} and voluntary prisoner’s dilemma game~\cite{sharma2023small}. Compared with cooperative agents, defective agents can improve cooperation in the snowdrift game among well-mixed populations~\cite{guo2023facilitating}. For the voluntary prisoner’s dilemma game, the loner agent has negligible influence in the well-mixed network but promotes cooperation in the lattice network~\cite{sharma2023small}. Besides cooperative behavior, it is also witnessed that prosocial punishment agents can establish social punishment among normal humans in well-mixed and lattice populations, whereas anti-social punishment agents cannot~\cite{shen2024prosocial}. It further extends the research on the emergence of prosocial behaviors in hybrid systems, showing that it is crucial to consider factors such as the agents' strategies as well as the social dilemmas of interaction environment.
 
 So far, scholars have highly focused on objective performance metrics, which are used to assess the potential ability of an agent to cooperate with humans, while blurring human subjective interaction preferences for machine agents~\cite{mckee2024warmth}. Based on such a social reality, Google Duplex can realistically imitate human language, thereby hiding their true nature and pretending to be a human~\cite{hern2018google}. Such deception has raised concerns and calls for the identity of machine agents should be transparent when interacting with humans~\cite{harwell2018google}. Recently, behavioral experiments have confirmed that in repeated prisoner's dilemma games between humans and agents, agents are more likely to stimulate cooperative behavior than humans, but this conclusion is no longer applicable once the true identities of agents are revealed. Even when the agents showed a more cooperative attitude over time, it did not change the participants' bias against the agents~\cite{ishowo2019behavioural}. It implies that humans have inherent interaction preferences for partners with different identities~\cite{oliveira2021towards}. Therefore, in human-agent hybrid systems with human-human interaction, agent-agent interaction, and human-agent interaction, which naturally raises people's questions about how individuals with different identities choose interaction partners. And how do individuals' interaction preferences for partners with different identities affect the evolution of cooperative behavior in hybrid systems?

\subsection{Contributions and organization}
 In order to understand these puzzles, this paper proposes a human-agent hybrid prisoner's dilemma game system under the framework of evolutionary game. In spatial networks, the most significant difference between agents and humans is the flexibility of decision-making behavior. Specifically, for the agent population, from the perspective of security, they follow traditional node dynamics, that is, adopting the same strategy to interact with different neighbors. However, different from the traditional committed agents, we endow agents with imitation ability, allowing them to update their strategies after each round of interaction with opponents, rather than maintaining a fixed strategy all the time. Compared with agents, considering that humans have stronger adaptive ability, on one hand, human group in the network are given more flexible decision-making freedom, allowing them to adopt different strategies to interact with different neighbors at the same time~\cite{wardil2010distinguishing,su2017evolutionary,jia2020evolutionary}, on the other hand, in addition to the two interacting parties, human behavior may be influenced by third individuals and pass on this influence to their partners, so as to more truly reflect human decision-making characteristics. Moreover, we introduce individuals' subjective interaction preferences for different groups, involving the interaction preferences between homogenous groups (i.e. human-human interaction,  and agent-agent interaction) and heterogeneous groups (i.e. human-agent interaction) respectively, and explore the evolution of cooperation in the hybrid system. The main contributions of this paper are as follows:

\begin{itemize}
\item This paper proposes a dynamic decision model based on individual's subjective interaction preferences, which provides insights into the evolution of cooperation in human-agent hybrid systems from the perspective of evolutionary game theory.
\item This paper reveals that individuals' asymmetric interaction preferences for different groups do help them perform better in complex interaction environments, especially stronger interaction preferences for heterogeneous groups. 
\item This paper shows that individuals with flexible decision mode can alleviate interaction dilemma, and increasing the agent's interaction preference for human can stimulate its own cooperative behavior.
\end{itemize}

The remainder of the paper is arranged as follows: In Section 2, we show the basic model of this work, including the network structure of the hybrid human-agent system, and the strategy learning rules for both agents and humans. In section 3, we present the results of numerical simulation. In Section 4, we propose an extended model with differentiated recognition ability, further comprehensively analyze the influence of interaction preference, recognition ability, cost and other factors on cooperation, and verify the robustness of individual subjective preference on the influence of cooperation evolution. Finally, we summarize the paper and discuss the future work.

\section{Model}
\subsection{Game environment of human-agent hybrid system}
We divide the groups in the spatial network into two types according to individuals' decision-making freedom, namely, agents who follow strict decisions and humans who follow free decisions. Agents and humans are evenly distributed in a lattice network with periodic boundaries with size $L=300$. Initially, all individuals randomly choose cooperation $(C)$ or defection $(D)$ strategies. Since the agent adopts the same strategy for all neighbors, and for any agent $x$, it strategy is denoted as $s_x=(s_{xy_1},s_{xy_2},\cdots,s_{xy_k})$, $s_{xy_1}=s_{xy_2}=\cdots=s_{xy_k}$, and $s_{xy_k}\in \left\{C,D\right\}$. Here $s_{xy_k}$ represents the focal individual strategy toward neighbor $y_k$. Therefore, the agent's action can be simply called cooperation or defection. While human group can adopt different strategies for different neighbors, and for any human individual $x$, its action can be expressed as a strategy set, $s_x=(s_{xy_1},s_{xy_2},\cdots,s_{xy_k})$, and $s_{xy_k}\in \left\{C,D\right\}$. In particular, strategies $s_{xy_i},i\in 1,2,\cdots,k$ are independent of each other. Then each individual interacts with $k (k=4)$ directly connected neighbors, and plays the weak prisoner's dilemma game. Where, the interaction between two cooperators will be rewarded $R=1$, and two defectors will be punished $P=0$, the cooperator interacts with the defector, the former player gets $S=0$, and the latter player gets the temptation to defect $T=b$ (it reflects the intensity of social dilemma). Each focal individual obtains cumulative income according to the following payoff matrix,

\begin{equation} 
  \bordermatrix{
      & C & D \cr
    C & R & S  \cr
    D & T & P  \cr
 }
\end{equation}

\begin{figure*}
\begin{center}
\includegraphics[scale=0.68]{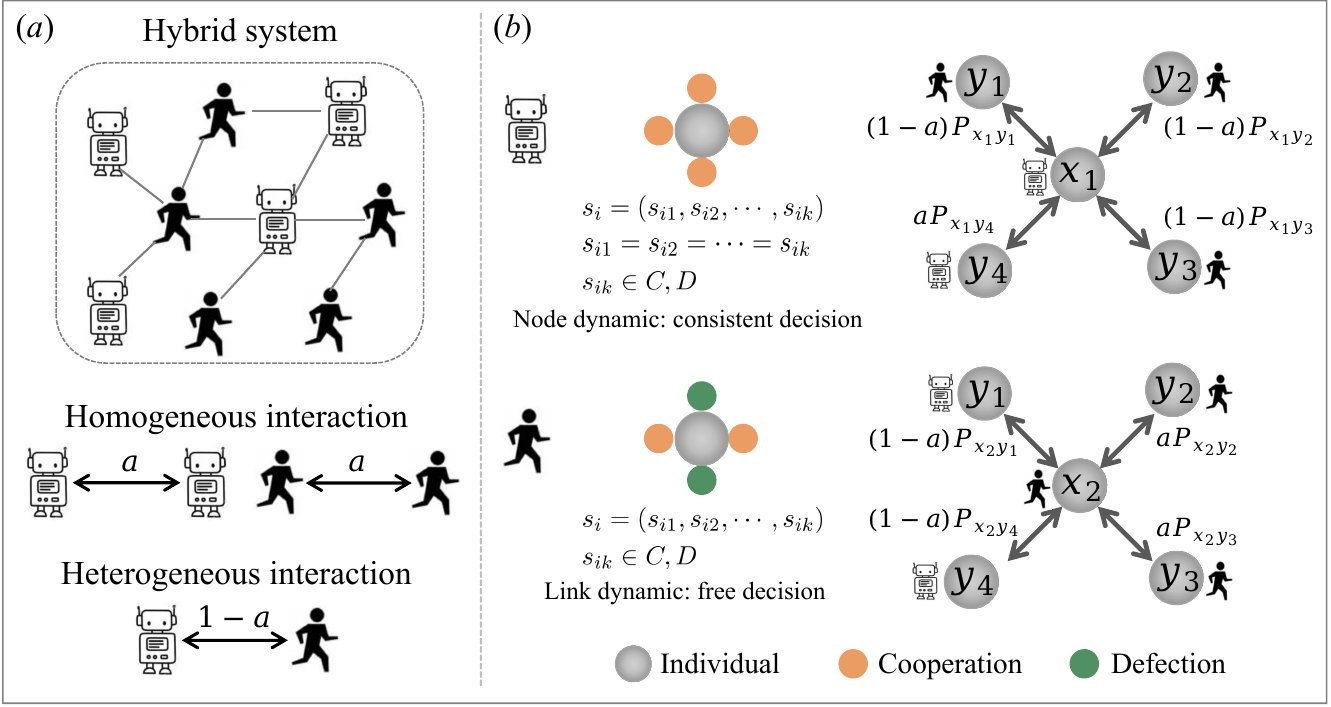}
\caption{\textbf{Diagram of human-agent hybrid system based on interaction preference.} Panel (\textit{a}) Two types of groups are randomly distributed in the spatial network, namely, agents group and human group. The focal individual interacts with both types of neighbors with differentiated preferences, and the interaction preference between homogeneous groups (i.e. human-human interaction, agent-agent interaction) is $\alpha$, and the interaction preference between heterogeneous groups (i.e. human-agent interaction) is $1-\alpha$. Panel (\textit{b}) During the game, agents follow the principle of taking consistent strategies to interact with their neighbors, called node dynamic, here $s_{ik}$ indicates the strategy of focal individual $i$ toward neighbor $k$, and $s_{i1}=s_{i2}=\cdots=s_{ik}$. Humans have flexible decision-making, allowing different strategies to interact with different neighbors, called link dynamic, here $s_{i1}$, $s_{i2}$ and $s_{ik}$ are independent of each other. Take focal agent $x_1$ and human $x_2$ as examples, their cumulative payoffs are $\alpha P_{{x_1}{y_4}}+(1-\alpha)(P_{{x_1}{y_1}}+P_{{x_1}{y_2}}+P_{{x_1}{y_3}})$, and $\alpha(P_{{x_2}{y_2}}+P_{{x_2}{y_3}})+(1-\alpha)(P_{{x_2}{y_1}}+P_{{x_2}{y_4}})$, respectively.}
\label{fig1} 
\end{center}      
\end{figure*}

\begin{figure*}
\begin{center}
\includegraphics[scale=0.63]{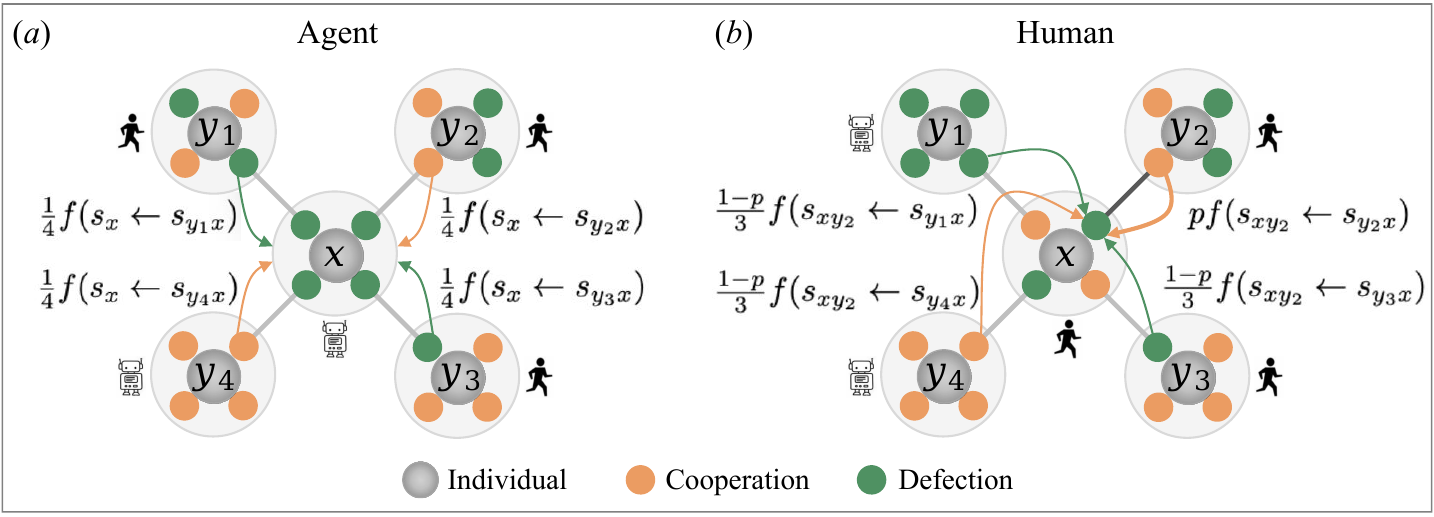}
\caption{\textbf{Diagram of strategy learning rules for two groups in human-agent hybrid system.} Panel (\textit{a}) The agent $x$ randomly selects a neighbor (so $p=0.25$), for example, neighbor $y_2$ is selected, and decides whether to imitate the selected neighbor $y_2$'s strategy $s_{y_2x}$ based on probability $f(s_{xy_2}\gets s_{y_2x})$ (that is, replace the agent's strategy $s_x$ with neighbor $y_2$'s strategy $s_{y_2x}$). Panel (\textit{b}) If human $x$ updates her strategy $s_{xy_2}$ towards neighbor $y_2$, she will select neighbor $y_2$ as the reference object with probability $p$, and imitate neighbor $y_2$'s strategy $s_{y_2x}$ according to probability $f(s_{xy_2}\gets s_{y_2x})$ (that is, replace human's strategy $s_{xy_2}$ with strategy $s_{y_2x}$). While randomly choose one from the other neighbors with probability $(1-p)/(k-1)$ (i.e. $y_3$) and imitate neighbor $y_3$’s strategy $s_{y_3x}$ according to probability $f(s_{xy_2}\gets s_{y_3x})$ (that is, replace human's strategy $s_{xy_2}$ with strategy $s_{y_3x}$).}
\label{fig2} 
\end{center}      
\end{figure*}

In a hybrid system, since any individual $x$ is likely to interact with both agent and human at the same time, the two groups with different identities will trigger the individual inherent subjective selection preference. In view of this, we introduce the parameter $\alpha$, which represents the interaction preference between homogeneous individuals, involving human-human interaction (HH) and agent-agent interaction (AA). While $1-a$ represents the interaction preference between heterogeneous individuals, like human-agent interaction (AH or HA), as Fig.1. Obviously, the focal individual's payoff from each neighbor not only depend on the strategies of the interacting parties, but also on the interaction preference of the focal individual toward the neighbor. Therefore, the cumulative payoff $\Phi_x$ of individual $x$ is expressed as:
\begin{equation}
\Phi_x=\sum_{w=0}^1\left[\sum_{i=1}^4(1-w)\alpha P_{xy_i}+\sum_{i=1}^4w(1-\alpha)P_{xy_i}\right]
\end{equation}
where, $P_{xy_i}$ means the payoff that individual $x$ gets by interacting with his neighbor $y_i$. What's more, the parameter $w$ serves as an indicator function, that is $w=0$ represents the interaction between homogeneous individuals, and $w=1$ means the interaction between heterogeneous individuals. In short, a larger parameter $\alpha$ implies that focal individuals are more inclined to interact with neighbors of the same type, whereas the opposite is true. Here $\alpha=0.5$ indicates that the focal individual interacts with groups with different identities equally.  

\begin{figure*}
\begin{center}
\includegraphics[scale=0.58]{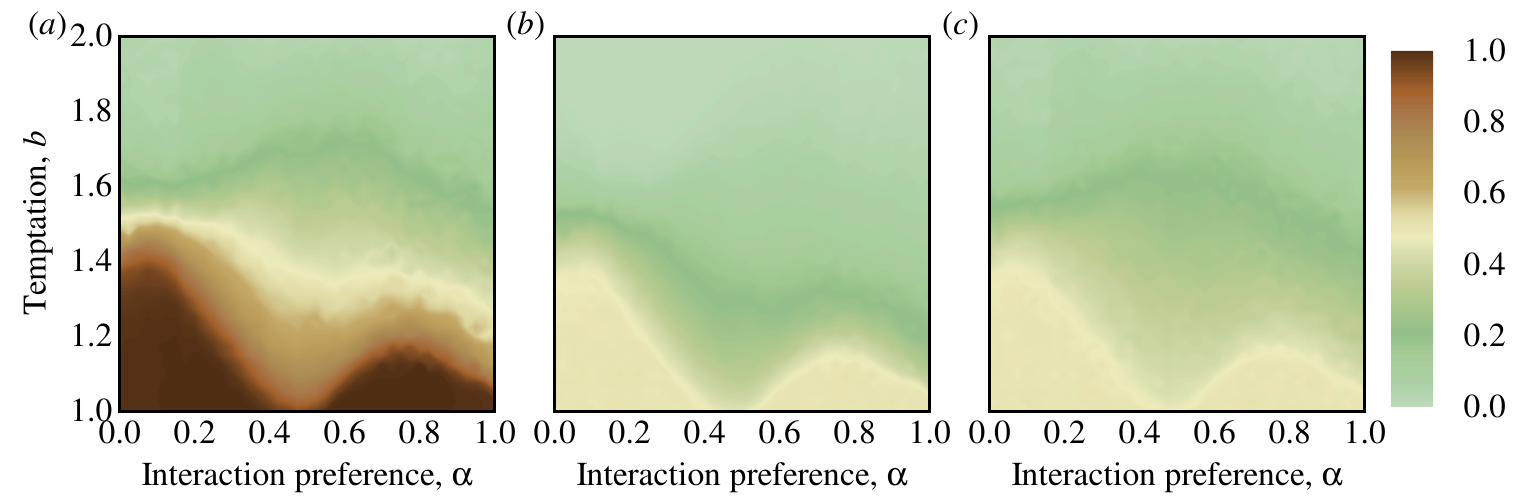}
\caption{\textbf{Asymmetric interaction preferences induce high-level cooperation.} The cooperation frequencies in steady state of total population, among the sub-population of agent group and human group are plotted as contours of interaction preference $\alpha$ and temptation $b$ from left to right. Panel (\textit{a}) Cooperative behavior of the system shows a double-hump pattern. Panels (\textit{b-c}) Agent and human groups show similar trends about cooperative behaviour. Individuals with flexible decision-making perform better in environments with greater social dilemmas.}
\label{fig3} 
\end{center}      
\end{figure*}

\subsection{Strategy learning rules}
In human-agent hybrid game, an asynchronous Monte Carlo iterative algorithm is used to update the strategy. Specifically, for each step of Monte Carlo iteration: first, an individual $x$ randomly selected interacts with neighbors and obtains cumulative payoff according to Eq.(2). Subsequently, the individual $x$ selects one of neighbors, i.e. $y_j$, as an object to imitate, and then learns the neighbor's strategy $s_{y_jx}$ with a probability $W$ measured by the Fermi function based on the payoff difference between the two individuals, as follows,
\begin{equation}
f(s_{x}\gets s_{{y_j}x})=\frac{1}{1+\exp{[-\beta({\Phi_{y_j}}-{\Phi_x})]}},
\end{equation}
where, $\Phi_x$ and $\Phi_{y_j}$ represent the cumulative payoffs of individual $x$ and selected neighbor $y_j$, respectively. Imitation strength $\beta (\beta \geq 0)$ measures how strongly individuals basing their decisions on payoff comparisons~\cite{sigmund2010social}. It means that the better selected neighbor $y_j$ performs, the more likely her strategy is to be imitated. Furthermore, for $\beta \rightarrow 0$ (weak imitation), the imitation is approximately random. For $\beta \rightarrow +\infty$ (strong imitation), the selected neighbor's strategy will always be imitated if her performs well, and never imitated otherwise. Following previous studies, here we set $\beta=10$~\cite{hauert2005game,chen2008promotion,sharma2023small} (we also explore how $\beta$ affects cooperation in more detail later). It is worth emphasizing that since individual $x$ only has the opportunity to imitate his own observed strategy, it is obvious that $s_{y_jx}$ refers to the strategy adopted by selected neighbor $y_j$ towards individual $x$, regardless of the strategies of selected neighbor $y_j$ towards his other neighbors.

Take individual $x$ updates his strategy against neighbor $y_i$ as an example, whether individual $x$ is an agent or a human, the imitation probability $W$ can be uniformly described as follows:
\begin{equation}
W=\left\{
\begin{array}{rcl}
p\frac{1}{1+exp[\beta(\Phi_x-\Phi_{y_j})]} & & {y_j\in\Omega,{y_j}={y_i}},\\
\frac{1-p}{k-1}\frac{1}{1+exp[\beta(\Phi_x-\Phi_{y_j})]} & & {y_j\in\Omega,{y_j}\neq{y_i}}.\\
\end{array} \right.
\end{equation}
Where parameter $\Omega$ represents the set of neighbors of individual $x$. 

It should be emphasized that since the two groups adopt different decision modes, they follow different learning rules, which are mainly reflected in how to choose a neighbor to imitate.

\begin{itemize}
\item \textbf{Agent's updating rule.} For an agent $x$, as shown in Fig.2, she will randomly select a neighbor $y_j$ (i.e. $y_2$), and learn the selected neighbor's strategy $s_{y_jx}$ (i.e. $s_{y_2x}$) with probability $W$. Thus, for each agent, $p=1/k$ (here $p=0.25$).

\item \textbf{Human's updating rule.} For a human $x$, from the perspective of direct reciprocity, she obeys a basic rule that if she updates her strategy towards which neighbor (i.e. $y_i$), then the neighbor $y_i$ is taken as the main object to imitate. Given that in the case of $p=1$, human $x$ completely dependent on directly related opponent (i.e. $y_i$), strategies can only be passed between two individuals who interact directly, causing the interaction relationship between humans to stop updating after it tends to be homogeneous (i.e. cooperation to cooperation, defection to defection)~\cite{su2017evolutionary}. Therefore, to avoid this dilemma, we set $p=0.99$. As shown in Fig.2, if the human $x$ updates the strategy towards neighbor $y_i$ (i.e. $y_2$), then the neighbor $y_j (y_j=y_i)$ (i.e. $y_2$) is selected as the imitation object with probability 0.99, while any other neighbor $y_j (y_j\neq y_i)$ (i.e. $y_1$, $y_3$, and $y_4$) is selected with probability $0.01/(k-1)$. Subsequently the human $x$ imitates the selected neighbor $y_j$'s strategy $s_{y_jx}$ with probability $W$. In particular, human $x$ updates her strategy towards each neighbor independently in the same way.
\end{itemize}

This model follows the Monte Carlo asynchronous update rule, that is, in a complete Monte Carlo algorithm, each individual is selected on average once to update his strategies for four neighbors. All results in steady state are measured by the average of the last $10^4$ of the total $5\times10^5$ steps. The parameters required for the model are shown in Table 1.

\begin{figure*}
\begin{center}
\includegraphics[scale=0.6]{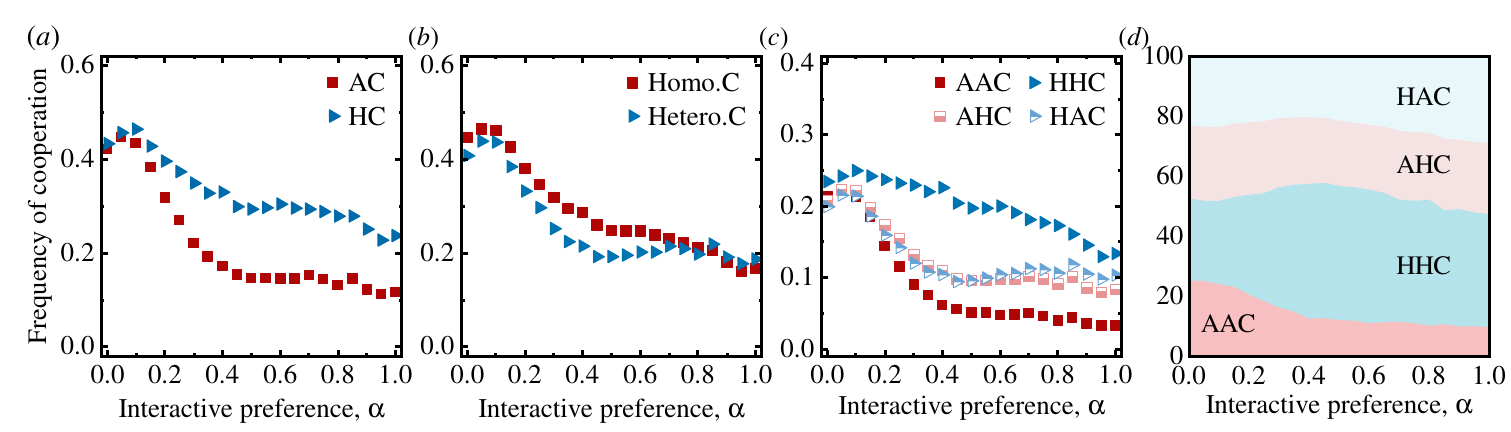}
\caption{\textbf{All interactions show higher cooperation when individuals have larger interaction preferences for heterogeneous groups ($\alpha$ is small)}. Panels (\textit{a-c}) The frequency of cooperation is plotted as a function of interaction preference $\alpha$ for individual types (agent and human), pairing types (homogeneous and heterogeneous), and pairing relationships (AA, HH, AH, and HA). Considering the directed interaction between players, AH and HA indicate different meanings, that is, AH indicates that the focal individual is an agent, and the paired interacting neighbor is a human, and so on. It shows that compared with agents, human groups are more likely to stimulate cooperative behavior in interactions. Panel (\textit{d}) The fraction of each pairing relationship in the cooperative population. All results are obtained for $b=1.4$.}
\label{fig4} 
\end{center}      
\end{figure*}

\begin{table}[ht]
\caption{Parameter values used in the model}
\label{Table 1}
\centering
\renewcommand{\arraystretch}{1.2}
\begin{tabular}{p{1.5cm}p{5.1cm} p{1.5cm}}
    \hline

     Parameter&Description & Value(s)\\
     \hline
     
     $N$&Network size& $9\times 10^4$\\
    \hline

     $\beta$&Imitation strength & 10\\
     \hline
     
     $b$&Temptation to defect& $[1,2]$\\
     \hline

    $\alpha$&Interaction preference & $[0,1]$\\
    \hline
     
    $u$&Density of identifiable individuals & $[0,1]$\\
    \hline

    $c$&Cost of obtaining information about the opponent's identity & $[0,1]$\\
    \hline
    
    \end{tabular}
\end{table}

\section{Results}
An asymmetric interaction preference for the two groups, that is, when the focal individual prefers to interact with one of the groups, is sufficient to induce a high level of cooperation (Fig.3). Cooperative behaviour shows a double-hump pattern with changes in the focal individual's interaction preference $\alpha$ towards the same type of neighbours. Specifically, the hump occurs when the interaction preference is 0.2 and 0.8, respectively. The first hump occurs in situations (e.g., $\alpha=0.2$ ) where focal individuals are more likely to interact with heterogeneous neighbors, while the second hump, in contrast, occurs in situations (e.g., $\alpha=0.8$ ) where focal individuals prefer to interact with homogeneous neighbours. In addition, the former shows higher cooperative behavior, and the trough of cooperation appears at $\alpha=0.5$. Moreover, the frequencies of cooperation of agent groups and human groups show similar trends (Fig.3 \textit{b},\textit{c}), but human groups are more likely to be triggered into highly cooperative behaviour, especially in environments with large social dilemmas.

In view of the complex and diverse interaction environment of population in the hybrid system, the cooperative behaviour of subgroups is analysed according to the type of individuals and interaction situation (Fig.4). From the perspective of individual type (Fig.4 \textit{a}), as the interaction preference of individuals changes from heterogeneous opponents to homogeneous opponents, the cooperative behavior of the two groups gradually declines. Overall human cooperative behavior (HC) is significantly higher than agent's cooperative behavior (AC). Subsequently, from the perspective of pairing relationship, it is found that the cooperative behavior of homogeneous interactions is higher than that of heterogeneous interactions (Fig.4 \textit{b}), which is mainly attributed to the high cooperation between human groups in homogeneous interactions. Specifically, by dividing the pairing relationship between individuals, it is observed that in the homogeneous interaction, the human-human (HH) pairing is more likely to form an interaction environment with high cooperation, while the agent-agent (AA) pairing falls into the dilemma of low cooperation. Cooperative behavior (i.e.,  human-agent interaction) in heterogeneous pairs is at a moderate level (Fig.4 \textit{c}). The distribution of cooperative behavior among the four paired interactions reflects that the interaction between humans dominates the cooperative behavior of the hybrid system (Fig.4 \textit{d}).

\begin{figure*}
\begin{center}
\includegraphics[scale=0.75]{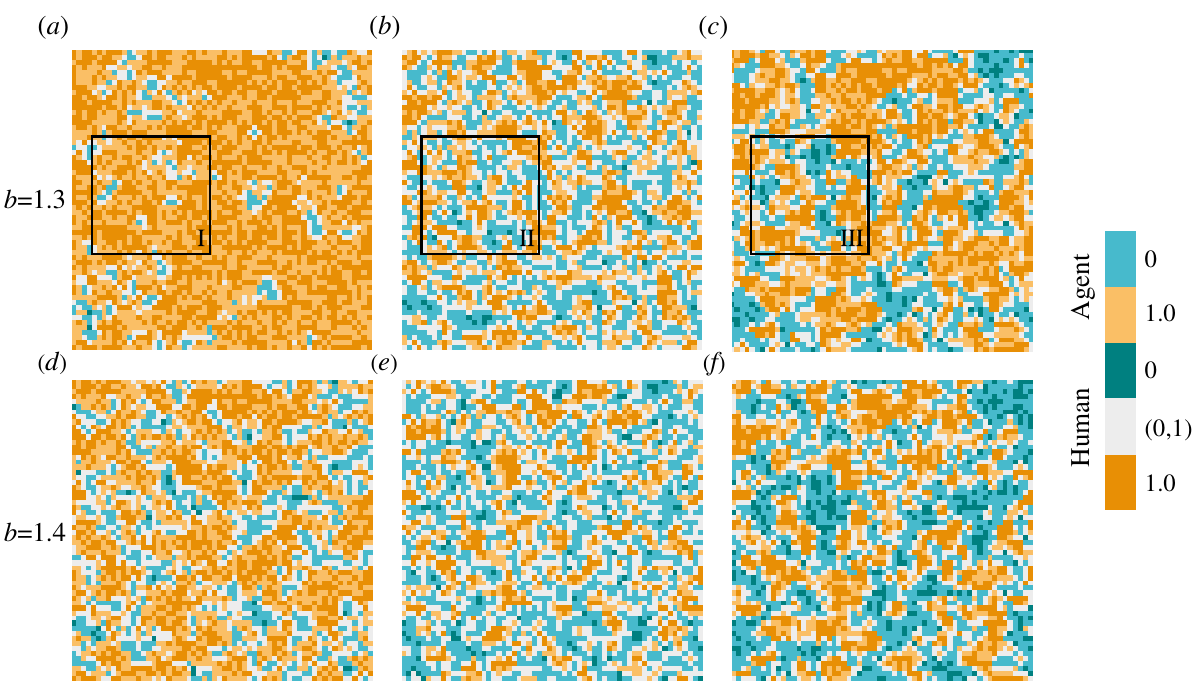}
\caption{\textbf{The asymmetric interaction preference of individuals for the two groups drives the hybrid system to form larger cooperative clusters.} Snapshots of the distribution of individual cooperation on grid network ($L=60$) in a steady state for different interaction preference $\alpha$, from left to right, the values of $\alpha$ are 0.2, 0.5, and 0.8, respectively. Results are obtained for $b=1.3$ (top), and $b=1.4$ (bottom). For each agent, the set of cooperation frequency is 0 (light cyan) and 1 (light orange), while for each human, the set of cooperation frequency is 0 (dark cyan), 0.25 (light gray), 0.5 (light gray), 0.75 (light gray), and 1 (dark orange). For ease of description, humans with cooperation frequencies of 0.25, 0.5, and 0.75 are uniformly denoted as $(0,1)$.}
\label{fig5} 
\end{center}      
\end{figure*}

In order to further explore the influence of interaction preference on individual cooperative behavior, we analyze the distribution of individual cooperation frequency from a micro perspective (Fig.5). The results show that individuals' asymmetric interaction preferences for different groups promote the hybrid system to form large clusters, just as the peak of the double hump of system cooperation appears in the cases of $\alpha=0.2$ and $\alpha=0.8$. Specifically, when individuals are more inclined to interact with heterogeneous opponents ($\alpha= 0.2$), the human group can stimulate the cooperative behavior of the agent group that interacts with it, and at the same time serve as a barrier to protect the cooperative cluster (Region I: The defective agents (light cyan) are surrounded by humans (light gray)). When individuals are more likely to interact with a homogeneous opponent ($\alpha=0.8$), humans surrounded by agents evolve into the same strategies as agents. For example, humans surrounded by cooperative agents (light orange) evolve into pure cooperators (dark orange) to obtain stable benefits, thus forming cooperative clusters, while humans surrounded by defective agents (light cyan) evolve into pure defectors (dark cyan) to avoid risks, thus forming defective clusters (Region III), which leads to an increase in the defective cluster, so that the second hump is lower than the first. When individuals treat two types of opponents equally ($\alpha=0.5$), due to a large number of small clusters are scattered in the spatial network, forcing humans make full use of their flexible decision-making forms to cooperate with cooperative clusters (orange) and defect with defective clusters (cyan), so that a large number of humans (light gray) with 0.25, 0.5, and 0.75 cooperation frequencies spontaneously appear in the system as protective bands (Region II). Comparing different social dilemmas, it is found that regardless of individual interaction preferences, humans act as a barrier to isolate cooperative clusters from defective clusters when necessary. This interesting behavioral trait is inherently derived from humans freely making decisions, and similar phenomena have been observed in previous research~\cite{jia2020evolutionary}.

\begin{figure*}
\begin{center}
\includegraphics[scale=0.68]{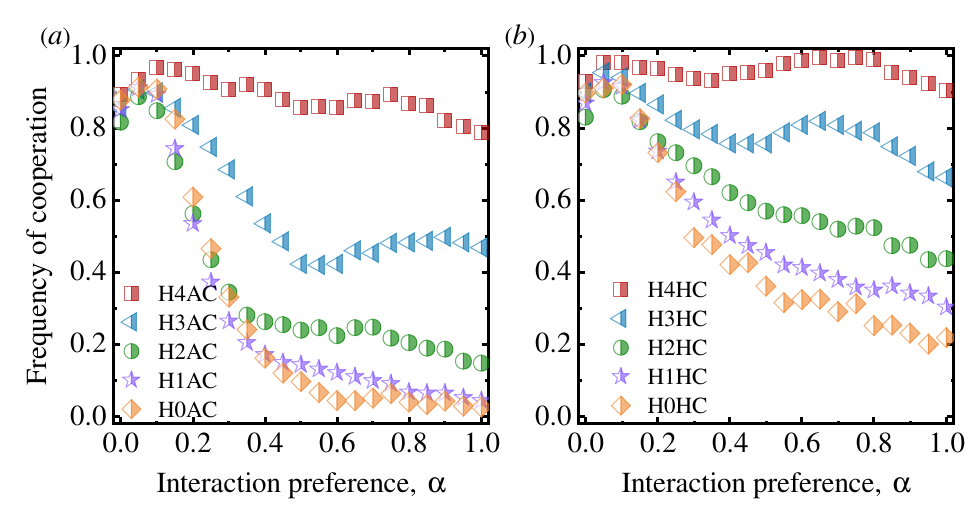}
\caption{\textbf{More humans in the neighborhood significantly stimulate individuals' cooperative behavior.} The average cooperative frequency of individuals is plotted against interaction preference $\alpha$, conditioned to the number of humans in their neighborhood for focal individuals are agents (left) and humans (right).  Thus as, H1AC denotes the cooperation rate of the agent whose one of the neighbors is human. H3HC denotes the cooperation rate of the human whose three of the neighbors are humans. Results are obtained for $b=1.4$. It is found that regardless of individual type and interaction preference, individuals surrounded by humans show stable cooperative behavior. In addition, both agents and humans surrounded by agents have to increase their interaction preferences with heterogeneous groups to alleviate social dilemmas.}
\label{fig6} 
\end{center}      
\end{figure*}

In view of the fact that the cooperative behavior of the agent and human groups shows a significant difference based on the type of neighbors, as shown in the spatial distribution of strategies (Fig.5). Therefore, we analyzed how the cooperation rate of the two types of groups changes with interaction preference $\alpha$ under different interaction environments, which is reflected in the number of humans in the neighbors (Fig.6). Regardless of the number of humans in the neighborhood, humans tend to cooperate more than agents. Moreover, regardless of whether it is agent or human, humans in the neighborhood are more likely to stimulate individual cooperative behavior. As observed, focal individuals surrounded by humans (H4AC and H4HC) exhibit stable and high levels of cooperation across a variety of interaction preference conditions. In particular, when agents are surrounded by agents (H0AC), they can reverse the dilemma of low cooperation by adjusting the interaction preference with different groups, such as reducing the interaction preference for agents. In short, the results show that in any case, the interaction between humans makes it easier for high cooperative behavior to flourish.

\begin{figure*}
\begin{center}
\includegraphics[scale=0.6]{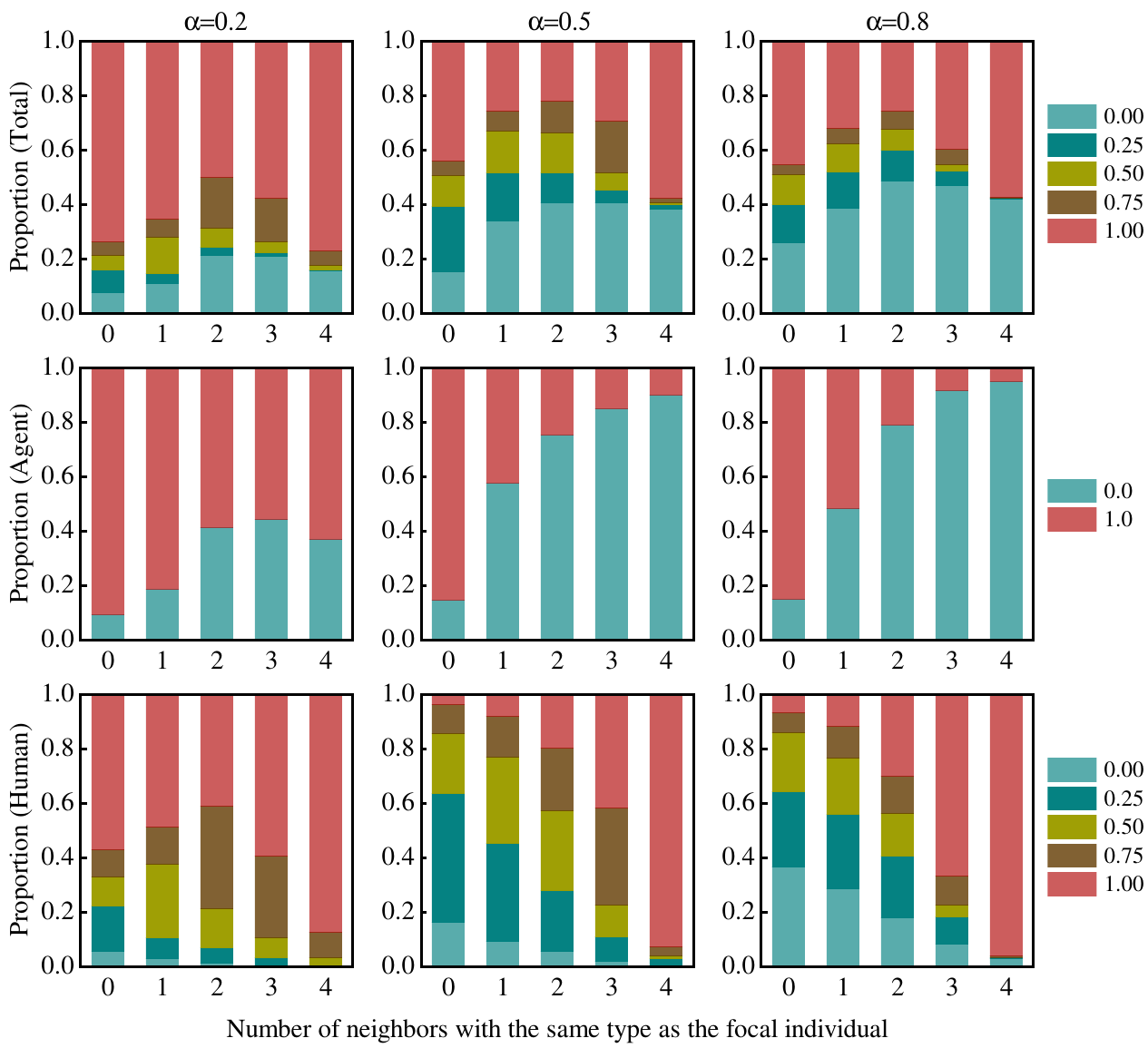}
\caption{\textbf{Agents enhance their cooperative behavior by increasing their interaction preferences with humans.} The distribution of individual cooperation frequencies of total population, subpopulation of agent group and human group depends on the number of neighbors with the same type of focal individual for different interaction preference $\alpha$, from left to right, the values of $\alpha$ are 0.2, 0.5, and 0.8, respectively. Results are obtained for $b=1.4$.}
\label{fig7} 
\end{center}      
\end{figure*}

\begin{figure*}
\begin{center}
\includegraphics[scale=0.57]{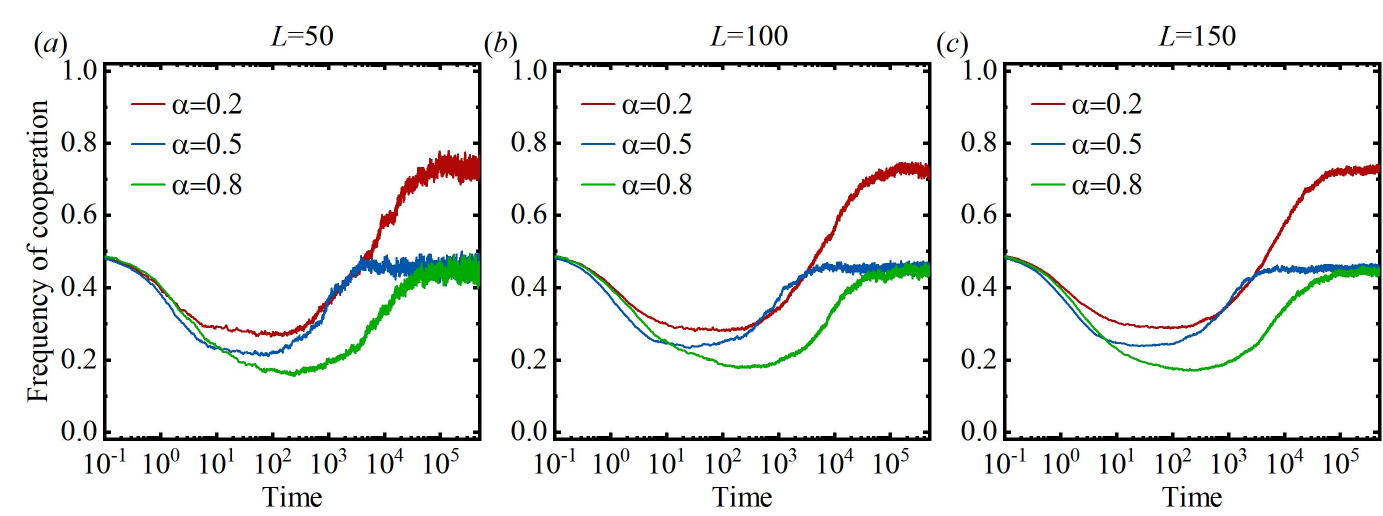}
\caption{\textbf{The size of the network group does not affect the evolution of cooperative behavior in the hybrid system.} The evolution of cooperative behavior in the hybrid system for different interaction preference $\alpha$, and $\alpha$ values are 0.2, 0.5, and 0.8, respectively. The size $L$ of the network is 50, 100, and 150, respectively. Results are obtained for $b=1.4$, and $\beta=10$.}
\label{fig8} 
\end{center}      
\end{figure*}

\begin{figure*}
\begin{center}
\includegraphics[scale=0.6]{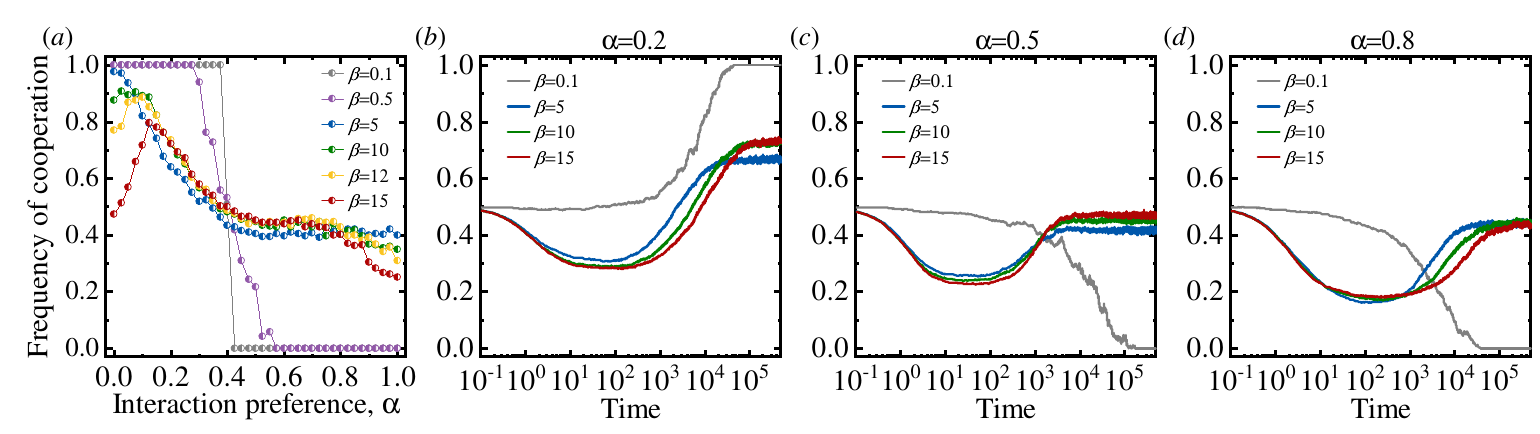}
\caption{\textbf{Asymmetric interaction preference, especially for heterogeneous groups, enhances the cooperative behavior of the system over a wide range of imitation strengths.} Panel (\textit{a}) The cooperation frequency of the system as a function of interaction preference $\alpha$ under different imitation strengths $\beta$. Panels (\textit{b-d}) The evolution of cooperative behavior in the hybrid system for different interaction preference $\alpha$, and $\alpha$ values are 0.2, 0.5, and 0.8, respectively. Results are obtained for $b=1.4$.}
\label{fig9} 
\end{center}      
\end{figure*}

Subsequently, in order to explore how the number of neighbors with the same type as the focal individual affects the cooperative behavior of individuals, we quantitatively analyzed the distribution of cooperation rates of individuals (Fig.7). Faced with different interaction preference $\alpha$, the distribution of individual cooperation rates is only slightly different from the perspective of the overall population. However, there is a significant difference in the distribution of individual cooperation rates between agent and human groups. Specifically, the cooperative behavior of agents is negatively correlated with the number of agents in the neighbors, and the negative correlation trend gradually becomes prominent with the interaction preference $\alpha$ increases. The behavior of human groups shows a completely opposite trend. The results suggest that agents in an environment with more agents in the neighborhood can improve their cooperative behavior by increasing their interaction preference for humans.

\begin{figure*}
\begin{center}
\includegraphics[scale=0.6]{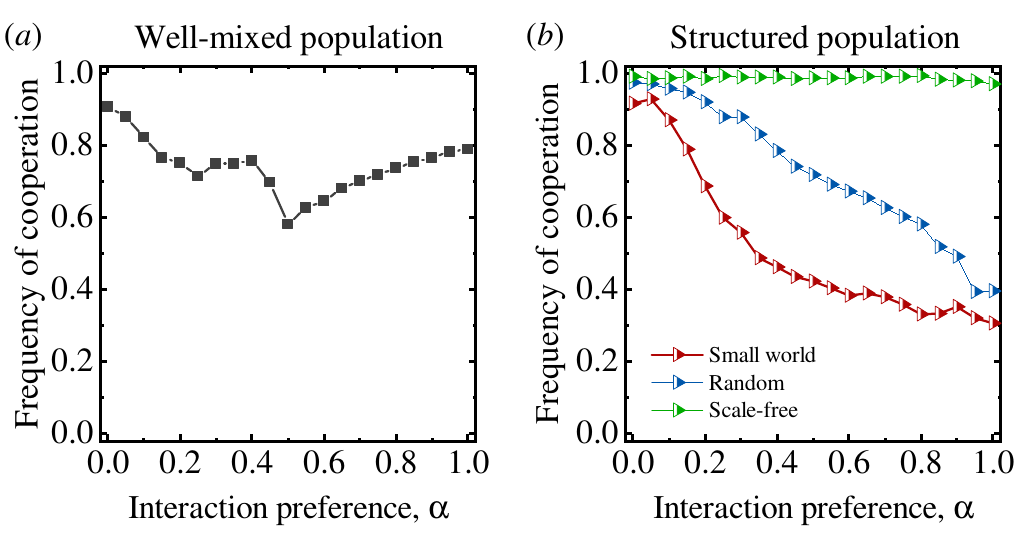}
\caption{\textbf{Asymmetric interaction preferences induce cooperative behavior in hybrid systems in well-mixed and structured populations.} Cooperative behavior in hybrid systems as a function of interaction preferences for different network populations. Results are obtained for $b=1.4$, and $\beta=10$.}
\label{fig10} 
\end{center}      
\end{figure*}

To verify the robustness of results, we further explore how network size $L$ and imitation strength $\beta$ affect the evolution of cooperation in hybrid system. It is found that for different interaction preferences, cooperative behavior in the hybrid system evolves independently of group size, except that we will obtain a more precise level of cooperation due to the fact that large network size reduces the fluctuation of cooperation frequency in the evolutionary steady state (Fig.8). In addition, the imitation strength greatly affects the evolution of cooperation (Fig.9). Specifically, for small imitation strengths (i.e. $\beta=0.1$ or $\beta=0.5$), the individuals' interaction preferences for heterogeneous groups (i.e. small $\alpha$) enable the system to achieve full cooperation, while interaction preferences for homogeneous groups (i.e. large $\alpha$) lead the system into the dilemma of total defection. With the increase of imitation strength, it is found that individuals' interaction preferences for any type of group can avoid the disappearance of cooperation. Although larger imitation strengths weaken cooperation in cases where individuals have strong interaction preferences for homogeneous or heterogeneous groups, it is confirmed that moderate interaction preference for heterogeneous groups can make the system maintain a high level of cooperation across a wide range of imitation strength.

Inspired by previous research, the effect of agents promoting cooperation in human-agent hybrid systems is related to population structure. Therefore, in the end, we provide the results of well-mixed population and other structured populations besides lattice, as shown in Fig.10. Here the population size is 1000 for well-mixed, and 1600 for structured populations such as small world (Rewiring probability is 0.1), random and scale-free networks. It is found that for well-mixed population, individuals with strong asymmetric interaction preferences can enhance cooperation, that is, individuals with the same attitude (i.e. $\alpha=0.5$) towards two groups lead to the lowest level of cooperation in the hybrid system. However, for structured populations, individuals' strong interaction preferences for heterogeneous groups are generally more likely to promote cooperation. The results for small world population are similar to lattice, but are affected by rewiring probability. Compared to small world population, the hybrid system presents a high level of cooperation in random and scale-free populations. Therefore, regardless of the population structure, asymmetric interaction preference is the key to induce cooperation in the hybrid system.

\section{Extended model for human-agent game with differentiated ability to obtain opponent's identity information}

\begin{figure*}
\centering
  \includegraphics[scale=0.6]{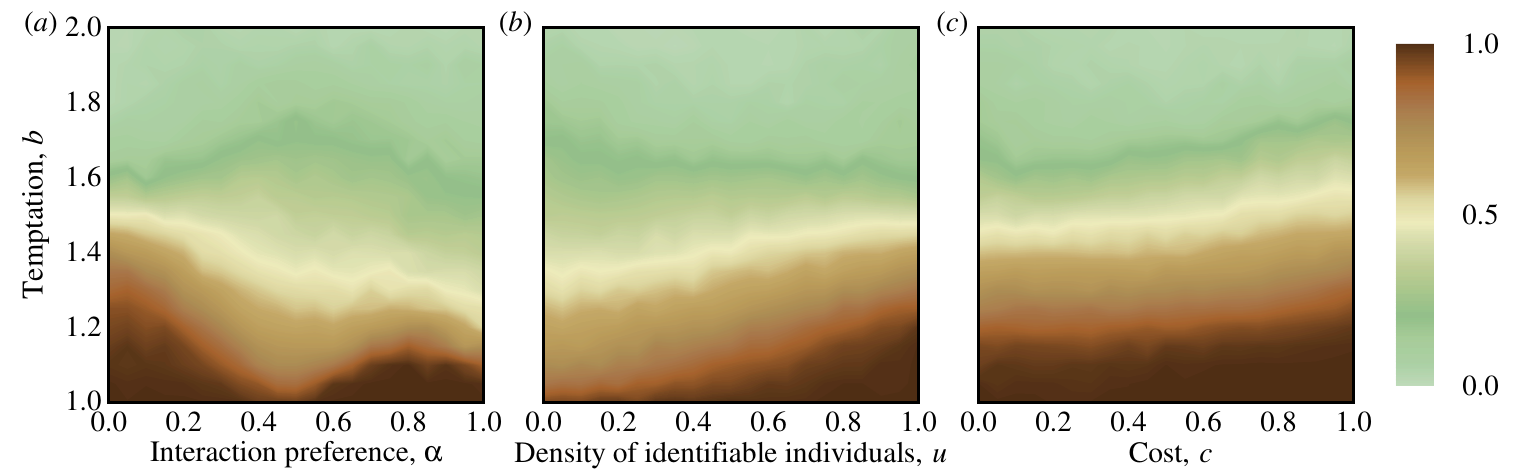}
\caption{\textbf{Contour plots of the cooperative behavior in $b-\alpha$, $b-u$, $b-c$ plane.} Panel (\textit{a}) The cooperative behavior of the system shows a double hump shape with the interaction preference $\alpha$, $c=0.2$, and $u=0.8$. Panel (\textit{b}) The cooperative behavior of the system increases significantly with the density of individual with identifiable ability, $c=0.2$, and $\alpha=0.2$. Panel (\textit{c}) High cost can appropriately enhance cooperative behavior in hybrid systems, $u=0.8$, and $\alpha=0.2$.} 
\label{fig11}       
\end{figure*}

\begin{figure*}
\centering
  \includegraphics[scale=0.6]{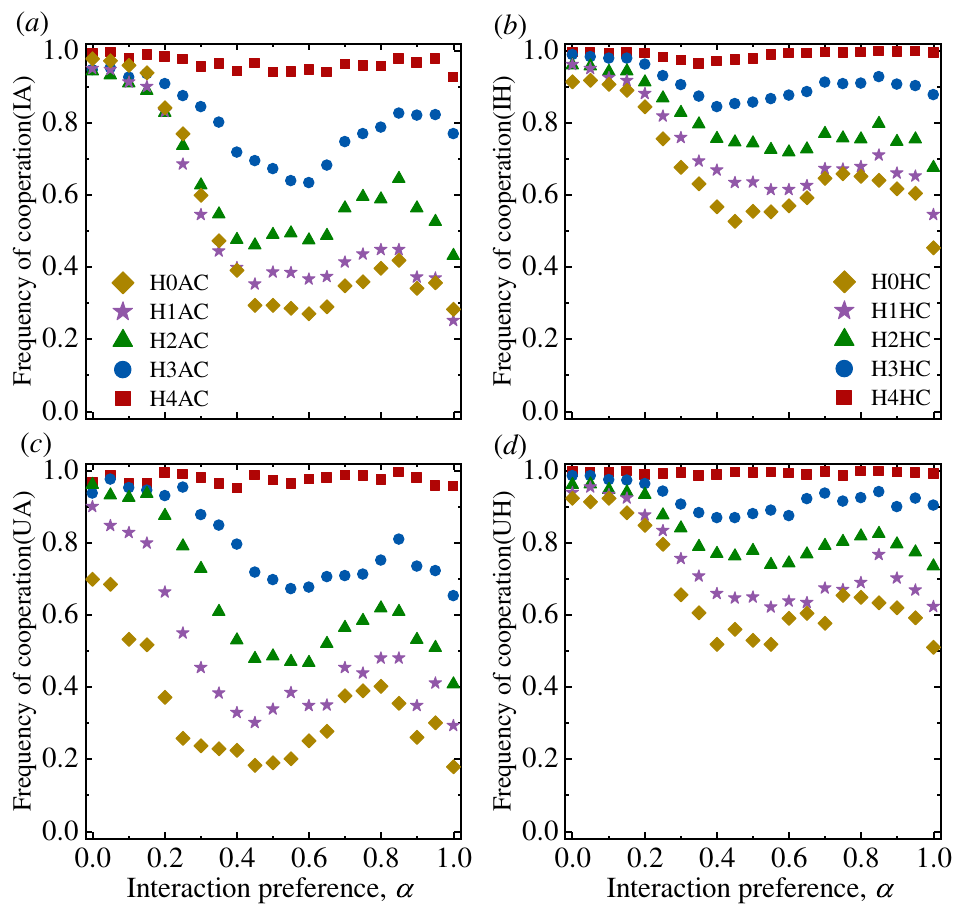}
\caption{\textbf{Agents with identifiable ability show higher cooperative behavior.} The cooperative behavior of individuals varies with the interaction preference $\alpha$ in the interaction environment with different numbers of humans in the neighborhood. Panels (\textit{a-d}) show the cooperative behavior of agents with identifiable ability (IA), humans with identifiable ability (IH), agents without identifiable ability (UA), and humans without identifiable ability (UH). All results are obtained for $c=0.2$, $b=1.2$, $u=0.8$.}
\label{fig12}       
\end{figure*}

\begin{figure*}
\centering
  \includegraphics[scale=0.6]{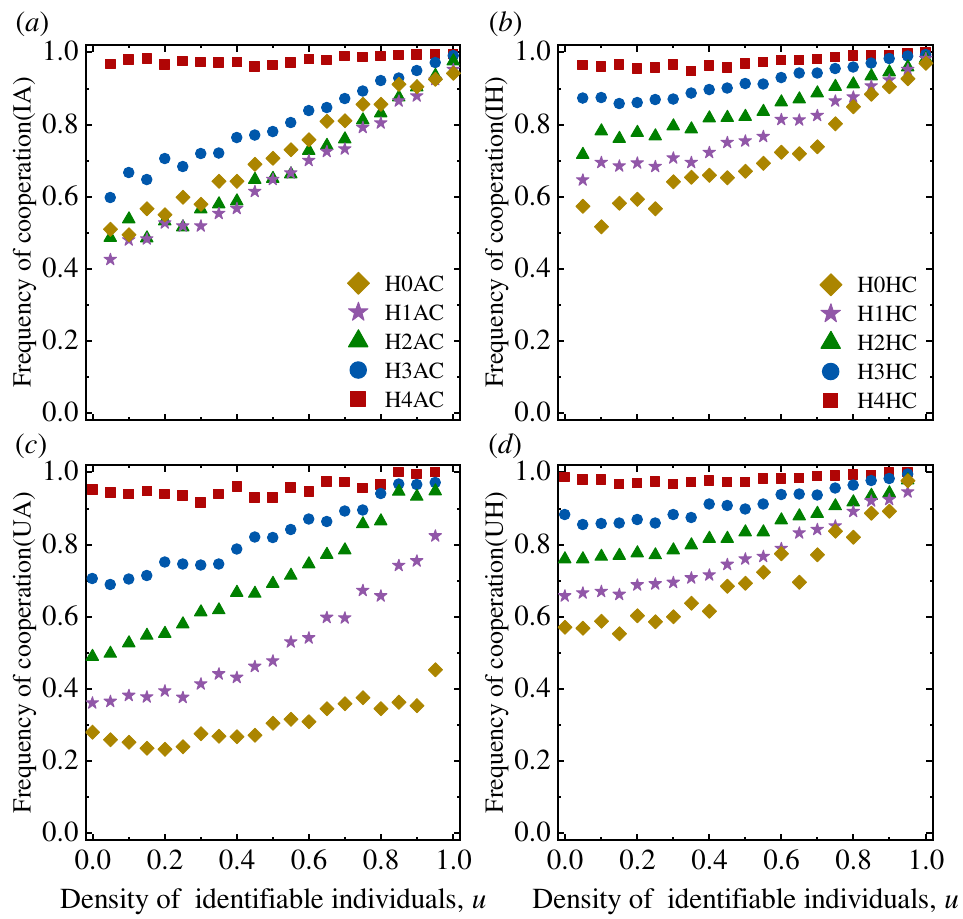}
\caption{\textbf{Increasing the density of individual with identifiable ability in the system can improve the cooperative behavior in various interactive environments, especially for agents with identifiable ability.} The cooperative behavior varies with the density of individual with identifying ability $u$ in the interaction environment with different numbers of humans in the neighborhood. Panels (\textit{a-d}) show the cooperative behavior of agents with identifiable ability (IA), humans with identifiable ability (IH), agents without identifiable ability (UA), and humans without identifiable ability (UH). All results are obtained for $c=0.2$, $b=1.2$, $\alpha=0.2$.}
\label{fig13}       
\end{figure*}

In the above model, it is assumed that all individuals’ identity information is transparent. That is, an individual interacts with a neighbor knowing whether the opponent is a human or a agent. Furthermore, considering that in the real environment, individuals have limited abilities and cannot obtain all information, people often have to pay for key information, such as opponent's identity information. Therefore, we improve the above model by introducing the cost factor, which not only considers the decision patterns of the two groups, but also pays attention to the individual's identifiable ability (that is, whether the individual pays a cost to obtain information about the opponent is a human or a agent). For convenience of description, we label the individual who obtain the opponent's identity information by paying a cost as ‘individual with identifiable ability’ , and conversely, the individual who do not obtain the opponent's identity information by paying the cost as ‘individual without identifiable ability’.

For the extended game model, which considers the individual's identifiable ability, agents and humans are evenly distributed on a regular lattice network with size $L=300$. The proportion of individuals with identifiable ability in the hybrid system is $u$. Therefore, the network interaction environment contains four types of subgroups, namely, agents with identifiable ability (IA), humans with identifiable ability (IH), agents without identifiable ability (UA), and humans without identifiable ability (UH). Individuals with identifiable ability and without identifiable ability respectively obtain cumulative payoff based on different payoff matrixes, where individuals without identifiable ability interact with their neighbours based on the payoff matrix Eq.(1), while the individuals with identifiable ability interact with their neighbors based on the payoff matrix as follows,

\begin{equation} 
  \bordermatrix{
      & C & D \cr
    C & 1-c & -c  \cr
    D & b-c & -c  \cr
 }
\end{equation}
where the parameter $c$ represents the cost that an individual with identifiable ability would have to pay to obtain each neighbor's identity information. There is no cost for individuals without identifiable ability. In particular, for individuals without identifiable ability, since such individuals cannot distinguish the types of opponents, there is no interaction preference when they interact with different types of neighbors, and the parameter $\alpha$ is fixed at 0.5. 

Through a large number of numerical simulations, we further analyze how interaction preference $\alpha$, the density of individual with identifiable ability $u$, and cost $c$ affect the evolution of cooperation in hybrid systems (Fig.11). The results show that the influence of interaction preference $\alpha$ on the cooperative behavior of the hybrid system is robust, which confirms that the asymmetric interaction preference of individuals for two groups promotes the system to achieve a high level of cooperation, especially the strong preference for heterogeneous groups. In addition, considering the cost and increasing the density of individuals with identifiable ability in the system can significantly improve the cooperative behavior.

Subsequently, whereas previous results have suggested that the cooperative behavior of individuals depends heavily on the interaction environment in which they are located. Therefore, we further analyze the cooperative behavior of humans and agents with and without identifiable ability in various neighborhood environments (Fig.12). It is observed that regardless of the flexibility of individual decision making, individuals with identifiable ability generally show higher levels of cooperation. Furthermore, even if individuals lack identifiable ability, flexible decision-making patterns help them perform well in various interaction environments, and even induce agents to present stable and high cooperation. The results in Fig.13 also confirm our findings.

\section{Conclusion and discussion}
Based on the inherent selection preferences of individuals in the real environment, this paper explores how the subjective interaction preferences of agents and humans for different groups affect cooperative behavior in a human-agent hybrid system. It is found that regardless of the population structure, compared with the previous research on the hybrid system without considering the interaction preference~\cite{su2017evolutionary,jia2020evolutionary}, asymmetric interaction preferences for two groups makes high cooperation appear in a wider range of social dilemmas, especially in the case of individuals' strong preference for heterogeneous groups. In addition, humans exhibit more stable prosocial behavior than agents, derived from the fact that humans can act as a barrier when necessary to separate cooperative clusters from defective clusters by inherently flexible decision making~\cite{jia2020evolutionary}. As observed, it is difficult to form a cooperative cluster in agent-centered interaction environment unless agents actively seeks interaction with humans, whereas humans surrounded by agents are more likely to evolve into pure cooperators. We further explored the case of incomplete information, where individuals need to pay the cost for opponent's identity information. On the one hand, it proves the robustness of the results in the case of transparent information. On the other hand, it is found that information transparency can effectively alleviate the interaction dilemma of agents, so that increasing the density of individuals with identifiable ability can enhance the cooperative behavior.

The current results imply that in order to enhance cooperative behavior in human-agent hybrid systems under the framework of evolutionary game, it is fruitful to improve the decision-making ability of both humans and agents. On the one hand, for humans, their decision-making behavior is depicted more closely to reality, as in this paper, in addition to the widely mentioned strategy imitation ability, we further endowing humans with flexible decision-making ability. On the other hand, for agents, their decision-making behavior is more anthropomorphic. As previous research has confirmed, committed agent improves human cooperation, but is constrained by the type of committed agent, game model, and network topology~\cite{guo2023facilitating,sharma2023small,shen2024prosocial}. In this paper, we give agents the ability of strategy imitation, so that they can learn the strategies of their well-performing neighbors through payoff comparison instead of always choosing the same strategy, so as to adapt to the complex interaction environment. Naturally, in the future research, we should further appropriately relax the agents' decision freedom, so that they can achieve flexible decision-making for each opponent in complex game environment like humans. Not limited to the anthropomorphic decision pattern, it is also effective to give agents anthropomorphic decision factors (i.e. interaction preference), allowing them to choose appropriate interaction partners based on different tasks or emotions. Thus, under the framework of evolutionary game, this study provides insights on how to improve and stabilize cooperative behavior in hybrid systems from the perspective of interaction preferences induced by decision patterns.

Moreover, it is not limited to the individual's interaction preferences for different groups. Recently, behavioral 
 experimental studies have confirmed that humans show a preference for one agent over the other, even among agents with the same performance~\cite{mckee2024warmth}. Therefore, in future work, we will explore the diversity of interaction preferences within and between groups, and further study the co-evolution of dynamic interaction preferences and cooperative behavior. Further, we will pay attention to the subjective preferences of different groups. For example, in homogeneous interaction pairs, human-human interaction preferences are different from agent-agent interaction preferences. Similarly, in heterogeneous interaction pairs, humans' preference for agents is different from agents' preference for humans. We strive to explore how to achieve better cooperation between humans and agents in hybrid systems from the local interactions between human-human, agent-agent, and human-agent.

\paragraph{Ethics} This work did not require ethical approval from a human subject or animal welfare committee.

\paragraph{Conflict of interest} We declare that we have no conflict of interest.

\paragraph{Use of AI} We declare that we have not used AI-assisted technologies in creating this article.

\paragraph{Authors’ Contributions} D.J.: Conceptualization, Data curation, Formal analysis, Investigation, Methodology, Software, Visualization, Writing-review $\&$ editing; X.D.: Software, Investigation, Writing-review $\&$ editing; J.X., P.T., Y.S. and Z.W.: Supervision, Resources.

\paragraph{Acknowledgments} This research was supported by the National Science Fund for Distinguished Young Scholars (No. 62025602), the National Science Fund for Excellent Young Scholars (No. 62222606), the National Natural Science Foundation of China (Nos. 11931015, U1803263, 81961138010 and 62076238), Fok Ying-Tong Education Foundation, China (No. 171105), Technological Innovation Team of Shaanxi Province (No. 2020TD-013), Fundamental Research Funds for the Central Universities (No. D5000211001), the Tencent Foundation and XPLORER PRIZE.

\bibliography{main}

\end{document}